\documentclass[showpacs,preprintnumbers,amsmath,amssymb,twocolumn]{revtex4}

\usepackage{graphicx}
\usepackage{dcolumn}
\usepackage{txfonts,bm}
\usepackage{hyperref}
\usepackage{color}

\begin{document}

\title{The meson-exchange model for the $\Lambda\bar{\Lambda}$ interaction}

\author{Lu Zhao$^{1,2}${\footnote{Email: zhaolu@ihep.ac.cn}}, Ning Li$^{3}$, Shi-Lin Zhu$^{3}$, Bing-Song Zou$^{1,4}${\footnote{Email: zoubs@itp.ac.cn}}}

\affiliation{1. Institute of High Energy Physics and  Theoretical
Physics Center for Science Facilities, CAS, Beijing 100049, China
\\2. University of Chinese Academy of Sciences, Beijing 100049, China
\\3. Department of Physics and State Key Laboratory of Nuclear Physics and Technology, Peking University, Beijing 100871, China
\\4. State Key Laboratory of Theoretical Physics, Institute of Theoretical Physics, CAS, Beijing 100190, China}

\begin{abstract}

In the present work, we apply the one-boson-exchange potential
(OBEP) model to investigate the possibility of $Y(2175)$ and
$\eta(2225)$ as bound states of $\Lambda\bar{\Lambda}(^3S_1)$ and
$\Lambda\bar{\Lambda}(^1S_0)$ respectively. We consider the
effective potential from the pseudoscalar $\eta$-exchange and
$\eta^{'}$-exchange, the scalar $\sigma$-exchange, and the vector
$\omega$-exchange and $\phi$-exchange. The $\eta$ and $\eta^{'}$
meson exchange potential is repulsive force for the state $^1S_0$
and attractive for $^3S_1$. The results depend very sensitively on
the cutoff parameter of the $\omega$-exchange ($\Lambda_{\omega}$)
and least sensitively on that of the $\phi$-exchange
($\Lambda_{\phi}$). Our result suggests the possible interpretation
of $Y(2175)$ and $\eta(2225)$ as the bound states of
$\Lambda\bar{\Lambda}(^3S_1)$ and $\Lambda\bar{\Lambda}(^1S_0)$
respectively.

\end{abstract}

 \maketitle{}

\section{INTRODUCTION}\label{introduction}

In 2005, the BABAR Collaboration announced the first observation of
$Y(2175)$ in the initial-state-radiation process $e^+ e^-
\rightarrow \gamma_{ISR}\phi(1020) f_0 (980)$, and the experimental
data indicated that it is a $J^{PC}=1^{--}$ resonance with mass
$m=2175\pm 10\pm 15$ MeV and width $\Gamma=58\pm16\pm20$
MeV~\cite{Aubert:2006bu}. Later, the BES Collaboration also observed
the similar structure in the decay of $J/\psi \rightarrow \eta \phi
f_0(980)$ with about $5\sigma$ significance~\cite{Ablikim:2007ab}.
Since both $Y(2175)$ and $Y(4260)$ are produced in $e^+ e^-$
annihilation and exhibit similar decay patterns, $Y(2175)$ might be
interpreted as an $s\bar{s}$ analogue of the $Y(4260)$, or as an
$s\bar{s}s\bar{s}$ state that decays predominantly to $\phi(1020)
f_0 (980)$~\cite{Aubert:2006bu}. So far, the interpretations of
$Y(2175)$ include $qqg$ hybrid~\cite{Ding:2006ya,Ding:2007pc},
tetraquark state~\cite{Wang:2006ri,Chen:2008ej,Drenska:2008gr},
excited $1^{--}$ $s\bar{s}$ state~\cite{Shen:2009zze} and resonance
state of $\phi
K\bar{K}$~~\cite{MartinezTorres:2008gy,GomezAvila:2007ru}. Besides,
there are also some other very interesting speculations on
$Y(2175)$~\cite{Zhu:2007wz,AlvarezRuso:2009xn,Yuan:2008br}.

The $\eta(2225)$ was first observed by the MARK-III collaboration in
the radiative decays $J/\psi \rightarrow \gamma \phi
\phi$~\cite{Bai:1990hk}. Its mass and width were measured to be
2220~MeV and 150~MeV respectively while its quantum numbers was
assigned to be $J^{PC}=0^{-+}$. Later, the BES Collaboration also
observed a signal around 2240~MeV from a high statistics study of
$J/\psi \to \gamma\phi\phi$ in the $\gamma K^+ K^- K^0_L K^0_L$
final state~\cite{Yuan:2008br}. In Ref.~\cite{Li:2008we}, the
authors investigated the strong decays of $3^1S_0$ and $4^1S_0$
within the the framework of the $^3P_0$ meson decay model and found
that the $\eta(2225)$ was very hard to interpreted as the
$3^1S_0~s\bar{s}$ state but a good candidate for $4^1S_0~s\bar{s}$
state.

Note that both $Y(2175)$ and $\eta(2225)$ are close to the threshold
of $\Lambda\bar{\Lambda}$. If these two states are not the
conventional $s\bar s$ excited state in the quark model, an
interesting interpretation is that they might be the loosely bound
states of $\Lambda\bar{\Lambda}$. If we only consider the S-wave
molecular states, $Y(2175)$ and $\eta(2225)$ should be assigned as
$\Lambda\bar{\Lambda}(^3S_1)$ and $\Lambda\bar{\Lambda}(^1S_0)$
respectively. Actually, more than thirty year ago, Dover, {\it et
al.} studied the bound states of $\Lambda\bar{\Lambda}$ with the
orbital angular quantum $L\geq 1$ within the meson-exchange model.
The interested readers can refer to Ref.~\cite{Dover:1977jk}.

In the present work, we apply the one-boson-exchange potential
(OBEP) model, which works very well in interpreting the deuteron, to
investigate the possibility of $Y(2175)$ and $\eta(2225)$ as bound
states of $\Lambda\bar{\Lambda}(^3S_1)$ and
$\Lambda\bar{\Lambda}(^1S_0)$ respectively. As an effective theory,
the one-boson-exchange potential model contains the long-range force
coming from the pion-exchange, the medium-range force coming from
the sigma-exchange and the short-range force coming from the heavier
vector rho/omega/phi-exchange. So far, lots of efforts have been
spent on the investigation of the possible bound states composed of
a pair of mesons or baryons within the one-boson-exchange potential
framework. In Ref.~\cite{Liu2009c}, the authors performed a
systematic study of the possible bound states composed of a pair of
heavy meson and heavy anti-meson within the one-boson-exchange
framework. In Ref.~\cite{Machleidt1987}, using the Bonn
meson-exchange model, the authors performed a detailed and
systematic study of the nucleon-nucleon interaction. The
one-boson-exchange potential model leads to an excellent description
of the deuteron data, $NN$ scattering phase shifts and many other
observables.

The paper is organized as follows. After the introduction, we
present the scattering amplitude in Section~\ref{amplitude} and the
effective potential in Section~\ref{potential}. The numerical
results are given in Section~\ref{numerical}. We discuss our results
in Section~\ref{summary}.

\underline{}

\section{Scattering Amplitude}\label{amplitude}

\begin{figure}[ht]
  \begin{center}
    \rotatebox{0}{\includegraphics*[width=0.4\textwidth]{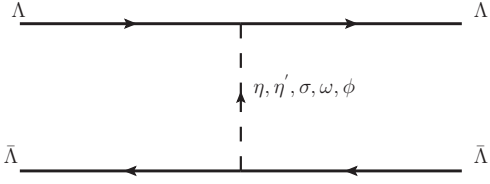}}
    \caption{The Feynman diagram at the tree level.\label{feynman}}
  \end{center}
\end{figure}

In the present work, we apply the Bonn meson-exchange model, which
works very well in the description of the deuteron, to calculate the
effective interaction potential of $\Lambda\bar{\Lambda}$.  In this
one-boson-exchange potential (OBEP) model, the long-range
$\pi$-exchange, the medium-range $\eta$-exchange and
$\sigma$-exchange, and the short-range $\omega$-exchange and
$\rho$-exchange combine to account for the interaction of the
loosely bound deuteron~\cite{Machleidt1987}. Given that the system
of $\Lambda\bar{\Lambda}$ is an isoscalar, the exchanged mesons
include $\eta$, $\eta^\prime$, $\sigma$, $\omega$. Besides, the
heavier $\phi$ should also account for the interaction of
$\Lambda\bar{\Lambda}$. The Feynman diagram at the tree level is
shown in Fig.~\ref{feynman}.

In our study, we first derive the baryon-baryon potential
$V_{\Lambda\Lambda}(r)$. Starting from $V_{\Lambda\Lambda}$, we
directly obtain $V_{\Lambda\bar{\Lambda}}$ by reversing the terms
corresponding to the exchange of a meson of odd G-parity, $G^i$,
i.e.,
\begin{eqnarray}
  V_{\Lambda\bar{\Lambda}}(r)=\sum_i(-1)^{G_i}V_{\Lambda\Lambda}^i(r).
\end{eqnarray}

The effective Lagrangian densities describing the
$\eta\Lambda\Lambda$, $\eta'\Lambda\Lambda$, $\sigma\Lambda\Lambda$
and $\omega\Lambda\Lambda$, $\phi\Lambda\Lambda$ vertices are

\begin{equation}
\mathcal{L}_{\eta\Lambda\Lambda}=-ig_{\eta \Lambda\Lambda}\bar\Psi
\gamma_{5}\Psi\eta,\label{eta-exchange}
\end{equation}

\begin{equation}
\mathcal{L}_{\eta'\Lambda\Lambda}=-ig_{\eta' \Lambda\Lambda}\bar\Psi
\gamma_{5}\Psi\eta',\label{etaprime-exchange}
\end{equation}

\begin{equation}
\mathcal{L}_{\sigma\Lambda\Lambda}=g_{\sigma
\Lambda\Lambda}\bar\Psi\sigma \Psi,\label{sigma-exchange}
\end{equation}

\begin{equation}
\mathcal{L}_{\omega\Lambda\Lambda}=-g_{\omega\Lambda\Lambda}\bar\Psi
\gamma_{\mu} \omega^{\mu}\Psi+
\frac{f_{\omega\Lambda\Lambda}}{2m_{\Lambda}}\bar\Psi\sigma_{\mu\nu}\Psi
\partial^{\mu}\omega^{\nu},\label{omega-exchange}
\end{equation}
and
\begin{equation}
\mathcal{L}_{\phi\Lambda\Lambda}=-g_{\phi\Lambda\Lambda}\bar\Psi
\gamma_{\mu}\phi^{\mu}\Psi+
\frac{f_{\phi\Lambda\Lambda}}{2m_{\Lambda}}\bar\Psi\sigma_{\mu\nu}\Psi
\partial^{\mu} \phi^{\nu}.\label{phi-exchange}
\end{equation}
In the above, $\Psi$ is the Dirac-spinor for the spin-$\frac{1}{2}$
particle of $\Lambda$. Actually, there only exits the vector form
for the $NN$ system with the $\omega$-exchange, $f_{\omega
NN}/g_{\omega NN}=0$~\cite{Machleidt1987}. Thus with the
$\mbox{SU}(3)-{flavor}$ symmetry, $f_{\omega \Lambda\Lambda}=f_{\phi
\Lambda\Lambda}=0$. Therefore, the Eqs~(\ref{omega-exchange}) and
(\ref{phi-exchange}) change into

\begin{equation}
\mathcal{L}_{\omega\Lambda\Lambda}=-g_{\omega\Lambda\Lambda}\bar\Psi
\gamma_{\mu} \omega^{\mu}\Psi,\label{omega-exchange-b}
\end{equation}
and
\begin{equation}
\mathcal{L}_{\phi\Lambda\Lambda}=-g_{\phi\Lambda\Lambda}\bar\Psi\gamma_{\mu}
\phi^{\mu}\Psi.\label{phi-exchange-b}
\end{equation}

With the Lagrangins given in
Eqs~(\ref{eta-exchange}-\ref{phi-exchange-b}), we can derive the
scattering amplitude of Fig.~\ref{feynman}. In our calculation we
adopt the Dirac spinor as
\begin{equation}
u(\vec{q},s)=\sqrt{\frac{E+M}{2M}}\begin{pmatrix}
                                     \chi_s \\
                                     \frac{\vec{\sigma}\cdot\vec{q}}{E+M}
                                   \end{pmatrix}\chi_s
\end{equation}
and
\begin{equation}
\bar u(\vec{q},s)\equiv
u^{\dag}(\vec{q},s)\gamma^0=\sqrt{\frac{E+M}{2M}}\begin{pmatrix}
                                     \chi^{\dag}_s  &&
                                      -\chi^{\dag}_s\frac{\vec{\sigma}\cdot\vec{q}}{E+M}
                                   \end{pmatrix}
\end{equation}

\begin{figure}[ht]
  \begin{center}
    \rotatebox{0}{\includegraphics*[width=0.4\textwidth]{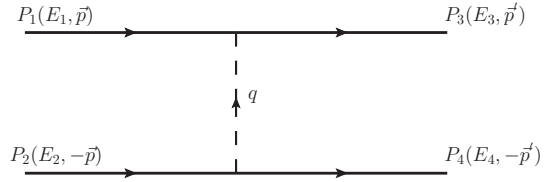}}
    \caption{ The four momentum for the $\Lambda\bar{\Lambda}$ system.
    \label{t-channel}}
  \end{center}
\end{figure}

In the center-of-mass frame, the initial four-momentums are
$P_1(E_1,\vec{p})$ and $P_2(E_2,-\vec{p})$ while the final
four-momentums are $P_3(E_1,\vec{p'})$ and $P_4(E_2,-\vec{p'})$, see
Fig.\ref{t-channel}. Thus the four-momentum of propagator is
\begin{equation}
q=P_3-P_1=P_2-P_4=(0,\vec{p'}-\vec{p})=(0,\vec{q})
\end{equation}
For the convenience of algebraic calculations, we make the following
momentum substitution,
\begin{equation}
\vec{q}=\vec{p'}-\vec{p}
\end{equation}
and
\begin{equation}
\vec{k}=\frac{1}{2}(\vec{p}+\vec{p'}).
\end{equation}

In our calculation, we make the nonrealistic approximation and keep
the terms up to order of $\frac{1}{m_{\Lambda}^2}$. The scattering
amplitudes are

\begin{eqnarray}
iM_{\eta}=&&g_{\eta\Lambda\Lambda}^2 \bar{u}_\Lambda \gamma_5
u_{\Lambda} \frac{i}{q^2-{m_\eta}^2}
\bar{u}_{\Lambda}\gamma_5u_{\Lambda} \nonumber\\
=&& i \frac{g_{\eta\Lambda\Lambda}^2}{\vec{q}^{2}+{m_\eta}^2}
\frac{(\vec{\sigma_1} \cdot \vec{q}) (\vec{\sigma_2} \cdot
\vec{q})}{4 {m_\Lambda}^2},\label{eta-amplitude}
\end{eqnarray}

\begin{eqnarray}
iM_{\sigma}=&&-g_{\sigma\Lambda\Lambda}^2\bar{u}_{\Lambda}u_{\Lambda}
\frac{i}{q^2-m_{\sigma}^2}
\bar{u}_{\Lambda}u_{\Lambda} \nonumber\\
=&&i\frac{g_{\sigma\Lambda\Lambda}^2}{\vec{q}^{2}+{m_\sigma}^2}
\left[1 - \frac{\vec{k}^{2}}{2 m_{\Lambda}^2} + \frac{\vec{q}^{2}}{8
m_{\Lambda}^2} + i\frac{\vec{S}\cdot(\vec{k}\times
\vec{q})}{2m_{\Lambda}^2}\right],\label{sigma-amplitude}
\end{eqnarray}
and
\begin{eqnarray}
iM_\omega=&&-g_{\omega\Lambda\Lambda}^2\bar{u}_{\Lambda}\gamma^{\mu}u_{\Lambda}
i\frac{-g_{\mu\nu}+\frac{q_{\mu}q_{\nu}}{m_{\omega}^2}}{q^2-m_{\omega}^2}\bar u_{\Lambda}\gamma^{\nu} u_{\Lambda}\nonumber\\
=&&i\frac{g_{\omega\Lambda\Lambda}^2}{\vec{q}^2+m_{\omega}^2}
\Bigg[1-\frac{\vec{q}^{2}}{8m_{\Lambda}^2} + \frac{3\vec{k}^{2}}{2
m_{\Lambda}^2}+
i\frac{3\vec{S}\cdot\left(\vec{k}\times\vec{q}\right)}{2m_{\Lambda}^2}\nonumber\\
&&-\frac{(\vec{\sigma_1}\cdot\vec{\sigma_2})\cdot\vec{q}^2}{4m_{\Lambda}^2}
+\frac{(\vec{\sigma}_1\cdot\vec{q})(\vec{\sigma}_2\cdot\vec{q})}{4m_{\Lambda}^2}\Bigg],
\label{omega-amplitude}
\end{eqnarray}
where $\vec{S}=\frac{1}{2}(\vec{\sigma}_1+\vec{\sigma}_2)$ is the
total spin of $\Lambda\bar{\Lambda}$. $iM_{\eta^{'}}$ is similar to
$iM_{\eta}$ while $iM_{\phi}$ is similar to $iM_{\omega}$. Making
the substitutions $g_{\eta\Lambda\Lambda}\to
g_{\eta^{'}\Lambda\Lambda}$ in Eq.~(\ref{eta-amplitude}) and
$g_{\omega\Lambda\Lambda}\to g_{\phi\Lambda\Lambda}$ in
Eq.~(\ref{omega-amplitude}), one can straightforwardly obtain
$iM_{\eta^{'}}$ and $iM_{\phi}$ respectively.


The coupling constants for the nucleon-nucleon-meson have been fixed
quite well by fitting the experimental data. In the present work, we
take the values of $g_{\alpha NN}$ from the Bonn meson-exchange
model~\cite{Machleidt1987}. The values of the coupling constants
$g_{\alpha\Lambda\Lambda}$ can be derived from $g_{\alpha NN}$
through the $\mbox{SU}(3)$-flavor symmetry. And they are

\begin{equation}
g_{\Lambda\Lambda\eta}=-\alpha\sqrt{\frac{4}{3}}
g_{NN\pi}(\cos\theta + \sin\theta),
\end{equation}
and
\begin{equation}
g_{\Lambda\Lambda\eta'}=\alpha\sqrt{\frac{4}{3}}
g_{NN\pi}(\cos\theta - \sin\theta),
\end{equation}
in which the quadratic Gell-Mann-Okubo mass formula is used and
\begin{equation}
\alpha\equiv D/(D+F)=0.6,
\end{equation}
\begin{eqnarray}
g_{\Lambda\Lambda\sigma}&=&\frac{2}{3} g_{NN\sigma}, \\
g_{\Lambda\Lambda\omega}&=&\frac{2}{3} g_{NN\omega},\\
g_{\Lambda\Lambda\phi}&=&\frac{1}{3} g_{NN\omega}.
\end{eqnarray}

The mass of $\Lambda$ is taken as $1115.7$ MeV from
PDG~\cite{Beringer:1900zz}. We summarize the numerical values of the
coupling constants and the masses of the exchanged mesons in
Table~\ref{coupling-mass}.

\begin{table}
\renewcommand{\arraystretch}{1.5}
\caption{The coupling constants $g_{\alpha\Lambda\Lambda}$, the
masses of the exchanged mesons taken from PDG~\cite{Beringer:1900zz}
and the cutoff parameters $\Lambda_\alpha$.}\label{coupling-mass}
\centering
\begin{tabular*}{8.5cm}{@{\extracolsep{\fill}}cccccc}
\hline \hline
$\alpha $    &  $\eta$     & $\eta'$    &   $\sigma$   & $\omega$  &  $\phi$\\
\hline
$m_\alpha$(MeV)&548.8      &957.7       &   550.0      &  782.6    &  1019.5\\
$\frac{g^{2}_{\alpha\Lambda\Lambda}}{4\pi}$ &4.473     &9.831      &3.459   &8.889       &2.222\\
$\Lambda_\alpha$(GeV)& 0.9 & 1.4        &0.9           &1.1        &1.5    \\
\hline\hline
\end{tabular*}
\end{table}

\section{Interactional Potential}\label{potential}

In the scattering theory of quantum mechanics, the relativistic
S-matrix has the form
\begin{equation}
\langle f | S | i \rangle = \delta_{fi} + i \langle f | T | i
\rangle = \delta_{fi} + (2\pi)^4\delta^4(p_f-p_i) i M_{fi},
\end{equation}
in which the T-matrix is the interaction part of the S-matrix and
$M_{ij}$ is defined as the invariant matrix element when extracting
the 4-momentum conservation of the T-matrix. The non-relativistic
S-matrix has the form
\begin{equation}
\langle f | S | i \rangle = \delta_{fi} - 2\pi \delta(E_f-E_i) i
V_{fi}.
\end{equation}

After considering both the relativistic normalization and
non-relativistic normalization, one gets the relationship between
the interaction potential $V_{fi}$ and the scattering amplitude
$M_{fi}$ in the momentum space.
\begin{equation}
V_{fi}=-\frac{M_{fi}}{\sqrt{\mathop\prod\limits_{f}2{E_f}
\mathop\prod\limits_{i}2{E_i}}}\approx-\frac{M_{fi}}
{\sqrt{\mathop\prod\limits_{f} 2{m_f}\mathop\prod\limits_{i}
2{m_i}}}.
\end{equation}

Considering the structure effect of the baryons, we introduce one
monopole form factor

\begin{equation}
F(q)=\frac{\Lambda^2-m_{ex}^2}{\Lambda^2-q^2}=\frac{\Lambda^2-m_{ex}^2}{{\Lambda}^2+\vec{q}^2},
\end{equation}
at each vertex. Here, $\Lambda$ is the cutoff parameter and $m_{ex}$
is the mass of the exchanged meson. To obtain the effective
potentials of the $\Lambda\bar{\Lambda}$ system, one needs to make
the following Fourier Transformation,

\begin{eqnarray}
  V(\vec{k},r)=\frac{1}{{(2\pi)}^3}\int d\vec{q}^3 e^{-i\vec{q}\cdot\vec{r}}V(\vec{q},\vec{k})F^2(\vec{q}),
\end{eqnarray}
and the following functions will be very helpful,

%
%
%
%
%

\begin{eqnarray}
\mathcal{F}_{1}&=&\mathcal{F}\left\{\frac{1}{\vec{q}^2+m^2}{\left(\frac{\Lambda^2-m^2}{{\Lambda}^2+\vec{q}^2}\right)}^2 \right\} \nonumber\\
&=& mY(mr)-\Lambda Y(\Lambda
r)-\left(\Lambda^2-m^2\right)\frac{e^{-\Lambda r}}{2
\Lambda},\label{transform-1}
\end{eqnarray}

\begin{eqnarray}
\mathcal{F}_{2}&=&\mathcal{F}\left\{{\frac{\vec{q}^2}{\vec{q}^2+m^2}}{\left(\frac{\Lambda^2-m^2}{{\Lambda}^2+\vec{q}^2}\right)}^2\right\}\\
&=& -m^3Y(mr)+ m^2\Lambda Y(\Lambda r) +
\left(\Lambda^2-m^2\right)\frac{\Lambda e^{-\Lambda
r}}{2},\label{transform-2}
\end{eqnarray}

\begin{eqnarray}
\mathcal{F}_{3} = && \mathcal{F}\left\{
\frac{(\vec{\sigma}_1\cdot\vec{q})(\vec{\sigma}_2\cdot\vec{q})}{\vec{q}^2+m^2}
\left(\frac{\Lambda^2-m^2}{{\Lambda}^2+\vec{q}^2}\right)^2\right\}\nonumber\\
=&&\frac{1}{3}\vec{\sigma}_1\cdot\vec{\sigma}_2\left[m^2\Lambda
Y(\Lambda r)-m^3 Y(mr)
+\left(\Lambda^2-m^2\right)\Lambda\frac{e^{-\Lambda r}}{2}\right]\nonumber\\
&&+\frac{1}{3}S_{12}\Bigg[-m^3 Z(mr)+ \Lambda^3 Z(\Lambda r)\nonumber \\
&&+(\Lambda^2-m^2)(1+\Lambda r)
\frac{\Lambda}{2}Y(\Lambda r)\Bigg]\nonumber\\
=&&(\vec{\sigma}_1\cdot\vec{\sigma}_2)\mathcal{F}_{3a} +
S_{12}\mathcal{F}_{3b},\label{transform-3}
\end{eqnarray}

\begin{eqnarray}
\mathcal{F}_{4}&=&\mathcal{F}\left\{{\frac{\vec{k}^2}{\vec{q}^2+m^2}}\left(\frac{\Lambda^2-m^2}{{\Lambda}^2+\vec{q}^2}\right)^2\right\}\nonumber\\
&=&\frac{m^3}{4}Y(Mr)-\frac{\Lambda^3}{4}Y(\Lambda
r)-\frac{\Lambda^2-m^2}{4}
\left(\frac{\Lambda r}{2}-1\right)\frac{e^{-\Lambda r}}{r} \nonumber\\
&~&-\frac{1}{2}\left\{\nabla^2,mY(mr)-\Lambda Y(\Lambda r)-\frac{\Lambda^2-m^2}{2}\frac{e^{-\Lambda r}}{\Lambda}\right\}\nonumber\\
&=&\mathcal{F}_{4a}+\left\{\nabla^2,\mathcal{F}_{4b}\right\},\label{transform-4}
\end{eqnarray}
and
\begin{eqnarray}
\mathcal{F}_{5}&=&\mathcal{F}\left\{i\frac{\vec{S}\cdot(\vec{q}\times\vec{k})}{\vec{q}^2+m^2}
\left(\frac{\Lambda^2-m^2}{{\Lambda}^2+\vec{q}^2}\right)^2\right\}\nonumber\\
&=&\vec{S}\cdot\vec{L}\left[-m^3Z_{1}(mr)+\Lambda^3Z_{1}(\Lambda r)+ (\Lambda^2-m^2)\frac{\Lambda e^{-\Lambda r}}{2r}\right]\nonumber\\
&=&\vec{S}\cdot\vec{L}\mathcal{F}_{5a}.\label{transform-5}
\end{eqnarray}
In the above equations, the functions $Y(x)$, $Z(x)$ and $Z_1(x)$
are defined as

\begin{equation}
Y(x)=\frac{e^{-x}}{x},
\end{equation}
\begin{equation}
Z(x)=\left(1+\frac{3}{x}+\frac{3}{x^2}\right)Y(x)
\end{equation}
and
\begin{equation}
Z_{1}(x)=\left(\frac{1}{x}+\frac{1}{x^2}\right)Y(x).
\end{equation}

With the help of Eqs.~(\ref{transform-1}-\ref{transform-5}), one can
easily write the effective potential of the system
$\Lambda\bar{\Lambda}$ as

\begin{eqnarray}
V_{\eta}(r)=-\frac{g_{\Lambda\Lambda\eta}^2}{4\pi}
\left(\frac{\vec{\sigma}_1\cdot\vec{\sigma}_2}{4
m_\Lambda^2}\mathcal{F}_{3a}+
\frac{1}{4m_{\Lambda}^2}S_{12}\mathcal{F}_{3b}\right),\label{eta-exchange-potential}
\end{eqnarray}

\begin{eqnarray}
V_{\sigma}(r)=\frac{g_{\Lambda\Lambda\sigma}^2}{4\pi}\left(-\mathcal{F}_{1}+
\frac{1}{2m_{\Lambda}^2}\mathcal{F}_{4a}-\frac{1}{8m_{\Lambda}^2}\mathcal{F}_{2}+
\frac{1}{2m_{\Lambda}^2}\vec{S}\cdot\vec{L}\mathcal{F}_{5a}\right),\label{sigma-exchange-potential}
\end{eqnarray}

\begin{eqnarray}
V_{\omega}(r)=&&\frac{g_{\Lambda\Lambda\omega}}{4\pi}\Bigg[-\mathcal{F}_{1}
-\frac{3}{2m_{\Lambda}^2}\mathcal{F}_{4a}+\frac{1}{8m_{\Lambda}^2}\mathcal{F}_{2} -\frac{3}{2m_{\Lambda}^2}\vec{S}\cdot\vec{L}\mathcal{F}_{5a}\nonumber\\
&&+
\frac{\vec{\sigma}_1\cdot\vec{\sigma_2}}{4{m_\Lambda}^2}(\mathcal{F}_{2}
- \mathcal{F}_{3a}) -
\frac{1}{4{m_\Lambda}^2}S_{12}\mathcal{F}_{3b}\Bigg],\label{omega-exchange-potential}
\end{eqnarray}
The effective potentials for the $\eta^{'}$-exchange and the
$\phi$-exchange are similar to $V_{\eta}$ and $V_{\omega}$
respectively. One can directly obtain them by making substitutions
$g_{\eta\Lambda\Lambda}\to g_{\eta^{'}\Lambda\Lambda}$ in
Eq.~\ref{eta-exchange-potential} and $g_{\omega \Lambda\Lambda}\to
g_{\phi\Lambda\Lambda}$ in Eq.~\ref{omega-exchange-potential}.

\begin{figure*}[ht]
  \begin{center}
  \rotatebox{0}{\includegraphics*[width=0.8\textwidth]{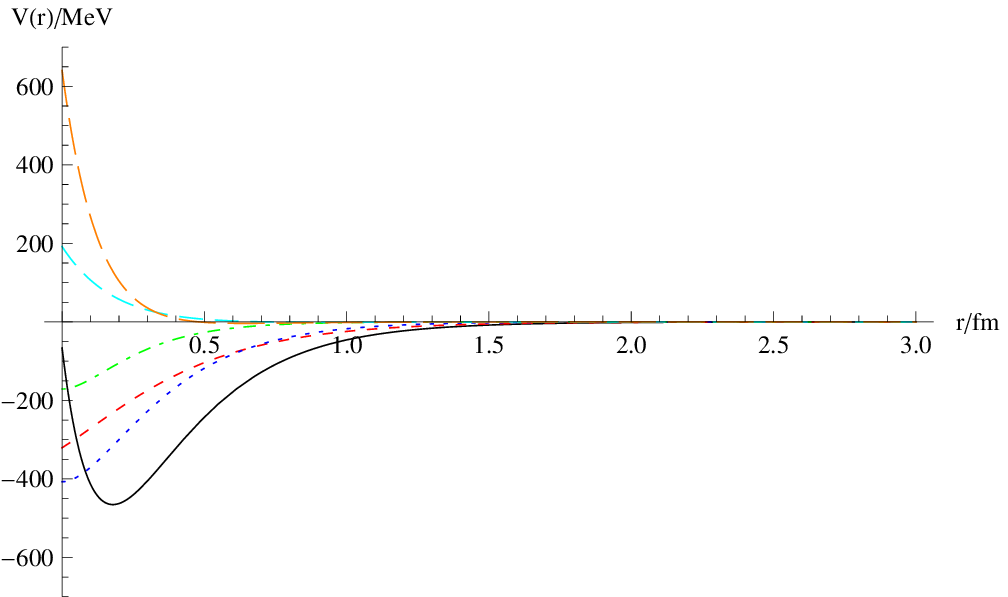}}
    \rotatebox{0}{\includegraphics*[width=0.8\textwidth]{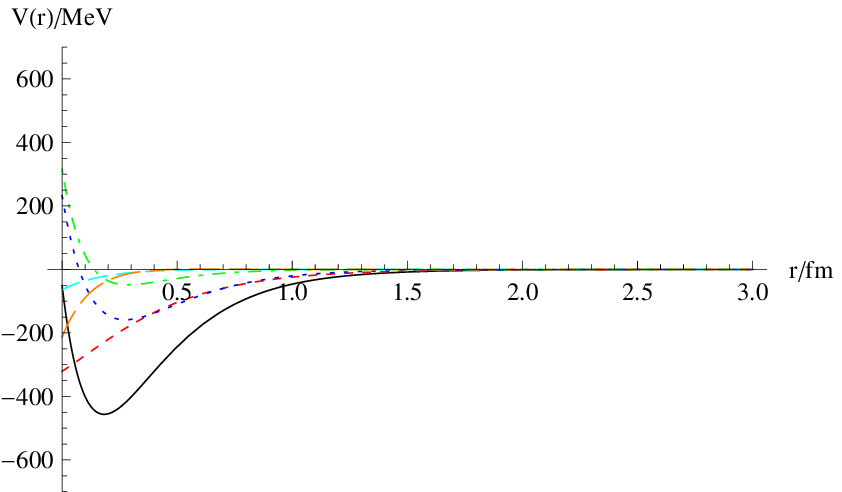}}
    \caption{(Color online) The effective potentials of the system $\Lambda\bar{\Lambda}$ with the parameters given in Table~\ref{coupling-mass}. The up one is for the state $^1S_0$ while the down one is for the state $^3S_1$. The solid line represents the total potential. The three dashing lines, short, medium and long, represent the contributions of the $\sigma$, $\eta$ and $\eta'$ exchanges respectively.
The contributions of the $\omega$-exchange is denoted by the dotted
line while that of the $\phi$-exchange is reflected by the
dot-dashed line. }
    \label{plot-potential}
  \end{center}
\end{figure*}

In order to make clear the specific roles of the exchanged mesons in
the effective potentials of the $\Lambda\bar{\Lambda}$ system, we
adopt a set of values of the cutoff parameters based on the mass of
the exchanged meson in Table~\ref{coupling-mass} and plot the
effective potential for states $^1S_0$ and $^3S_1$ in
Fig.~\ref{plot-potential}. From Fig \ref{plot-potential} we can see
that for the state $^1S_0$ both the $\eta$-exchange and the
$\eta^{'}$-exchange provide repulsive force while the
$\phi$-exchange, the $\omega$-exchange and the $\sigma$-exchange
provide attractive force. For the state $^3S_1$, the $\phi$-exchange
and the $\omega$-exchange provide repulsive force in the short range
but attractive force in the medium range while the
$\sigma$-exchange, the $\eta$-exchange and the $\eta^{'}$-exchange
provide attractive force.

\section{Numerical Results}\label{numerical}

The time-independent Schr\"odinger Equation is
\begin{equation}
\left[-\frac{\hbar^2}{2\mu}\nabla^{2}+V(\vec{r})-E\right]\Psi(\vec{r})=0.
\end{equation}

However, in our effective potential $V(\vec{r})$ there exist terms
related to $\nabla^2$. Thus for the convenience of algebraic
manipulation, we separate these momentum-dependent terms from the
potential and write the Shr\"odinger Equation in the form,

\begin{equation}
\left[-\frac{\hbar^2}{2\mu}\nabla^{2}-\frac{\hbar^2}{2\mu}\left(\nabla^2
\alpha(r)+\alpha(r)\nabla^2\right)+V_0(\vec{r})-E\right]\Psi(\vec{r})=0,
\end{equation}
in which $\alpha(r)$ has form
\begin{eqnarray}
\alpha(r)=(-2\mu)\left(\frac{g_{\sigma\Lambda\Lambda}^2}{4\pi}
\frac{1}{2}\mathcal{F}_{4b}-\frac{g_{\Lambda\Lambda\omega}^2}{4\pi}
\frac{3}{2}\mathcal{F}_{4b}\right)
\end{eqnarray}

In our calculation, we take the Laguerre polynomials as a complete
set of orthogonal basis to construct the radial wave function. The
normalized basis is

\begin{equation}
\chi_{nl}(r)=\sqrt{\frac{(2\lambda)^(2l+3) n!}{\Gamma(2l+3+n)}}r^l
e^{-\lambda r}L^{2l+2}_n (2\lambda r), n=1,2,3...,
\end{equation}
which satisfies

\begin{equation}
\int^\infty _0 \chi_{i}(r) \chi_{i}(r) r^2 dr=\delta_{ij}.
\end{equation}
Then the total wave function can be written as

\begin{equation}
\Psi(\vec{r})=\sum^{n-1}_{i=0}a_i \chi_{i 0}(r)|\Psi_S\rangle
\label{wavefunction}
\end{equation}
for the S-wave ($L=0$) of the $\Lambda\bar{\Lambda}$ system. With
this wave function, the spin-orbit interaction operator
$\vec{S}\cdot\vec{L}$ is 0. The spin-spin interaction operator
$\vec{\sigma}_1\cdot\vec{\sigma}_2$ is 1 for the state $^3S_1$ and
$-3$ for the state $^1S_0$. And, the tensor operator $\vec{S}_{12}$
is 0 for both states $^3S_1$ and $^1S_0$. Now with the initial state
$|i>$ and the final state $|f>$ the Hamiltonian of Schr\"odinger
Equation can be written in the following matrix form,

\begin{eqnarray}
H_{ij}=&&\int^\infty _0 a_i \chi_{i 0}(r) \Bigg[-\frac{\hbar^2}{2\mu}\left(1+\alpha(r)\right)\nabla^2 a_j \chi_{j 0}(r)\nonumber\\
&&-\frac{\hbar^2}{2\mu}\nabla^2 \left(\alpha(r)a_j \chi_{j 0}(r)\right)\nonumber\\
&&+ V (r,\vec{S}\cdot\vec{L}=0) a_j \chi_{j 0}(r)\Bigg] r^2 dr.
\end{eqnarray}

Digitalizing this matrix one can obtain the eigenvalue and the
eigenvector. If a negative eigenvalue is obtained, a bound state
exists. In our calculation, we apply a computational program which
is based on the variational method. We first vary the parameter
$\lambda$ to get the lowest value, then change the trial wave
function to reach a stable result.

For the cutoff parameter, we adopt the reasonable range as
$900\mbox{MeV}\sim2000\mbox{MeV}$ and follow the rule that the
heavier the mass is, the bigger the cutoff parameter is. Given that
both the $\eta$-exchange and the $\eta^{'}$-exchange provide
repulsive force for the state $^1S_0$ but attractive force for the
state $^3S_1$, we fix the cutoff parameter to be
$\Lambda_{\eta}=1900$ MeV for the $\eta$-exchange and
$\Lambda_{\eta^{'}}=2000$ MeV for the $\eta^{'}$-exchange to obtain
the largest differences between the states $^1S_0$ and $^3S_1$. For
the other cutoff parameters for the $\sigma$-exchange,
$\omega$-exchange and $\phi$-exchange, we tune them in the range
$900\mbox{MeV}\sim2000\mbox{MeV}$ to obtain the negative and stable
eigenvalue $E$ for the states $^1S_0$ and $^3S_1$. The numerical
results are shown in Table~\ref{eigenvalues}. Our results indicate
that when the cutoff parameters are adopted as
$\Lambda_{\sigma}\subset(900\sim1000~\mbox{MeV})$,
$\Lambda_{\omega}\subset(1050\sim1150~\mbox{MeV})$ and
$\Lambda_{\phi}\subset(1100\sim1200~\mbox{MeV})$, we obtain loosely
bound states of $\Lambda\bar{\Lambda}(^1S_0)$, with binding energy
being $-15.316\sim-73.517$ Mev. The binding of the state
$\Lambda\bar{\Lambda}(^3S_1)$ is much deeper, with binding energy
being $-59.045\sim-228.668$ MeV, see Table~\ref{eigenvalues}.

\begin{table}
\renewcommand{\arraystretch}{1.5}
\caption{Binding energies of the states
$\Lambda\bar{\Lambda}(^3S_1)$ and $\Lambda\bar{\Lambda}(^1S_0)$ with
different sets of $\Lambda_{\sigma}$, $\Lambda_{\omega}$ and
$\Lambda_{\phi}$. The cutoff parameter for the $\eta$-exchange and
$\eta^{'}$-exchange are fixed to be $\Lambda_{\eta}=1900$ MeV and
$\Lambda_{\eta^{'}}=2000$ MeV respectively.} \label{eigenvalues}
\centering
\begin{tabular*}{8.5cm}{@{\extracolsep{\fill}}ccccc}
\hline \hline
                        &                        &                      &\multicolumn{2}{c}{E (MeV)}\\
$\Lambda_{\sigma}$ (MeV)&$\Lambda_{\omega}$ (MeV)&$\Lambda_{\phi}$
(MeV)& $^{3}S_{1}$
& $^{1}S_{0}$\\
\hline 900                   & 1050                  &  1100
& $-59.054$
& $-15.316$\\
925                   & 1075                  & 1125            &
$-88.148$
& $-25.218$\\
 950                  & 1100                  & 1150            & $-124.938$
&$-38.011$\\
975                   & 1125                  & 1175            &
$-170.972$
&$-53.989$\\
1000                  & 1150                  & 1200            & $-228.668$ &$-73.517$\\
\hline\hline
\end{tabular*}
\end{table}

\begin{figure}[ht]
  \begin{center}
  \rotatebox{0}{\includegraphics*[width=0.47\textwidth]{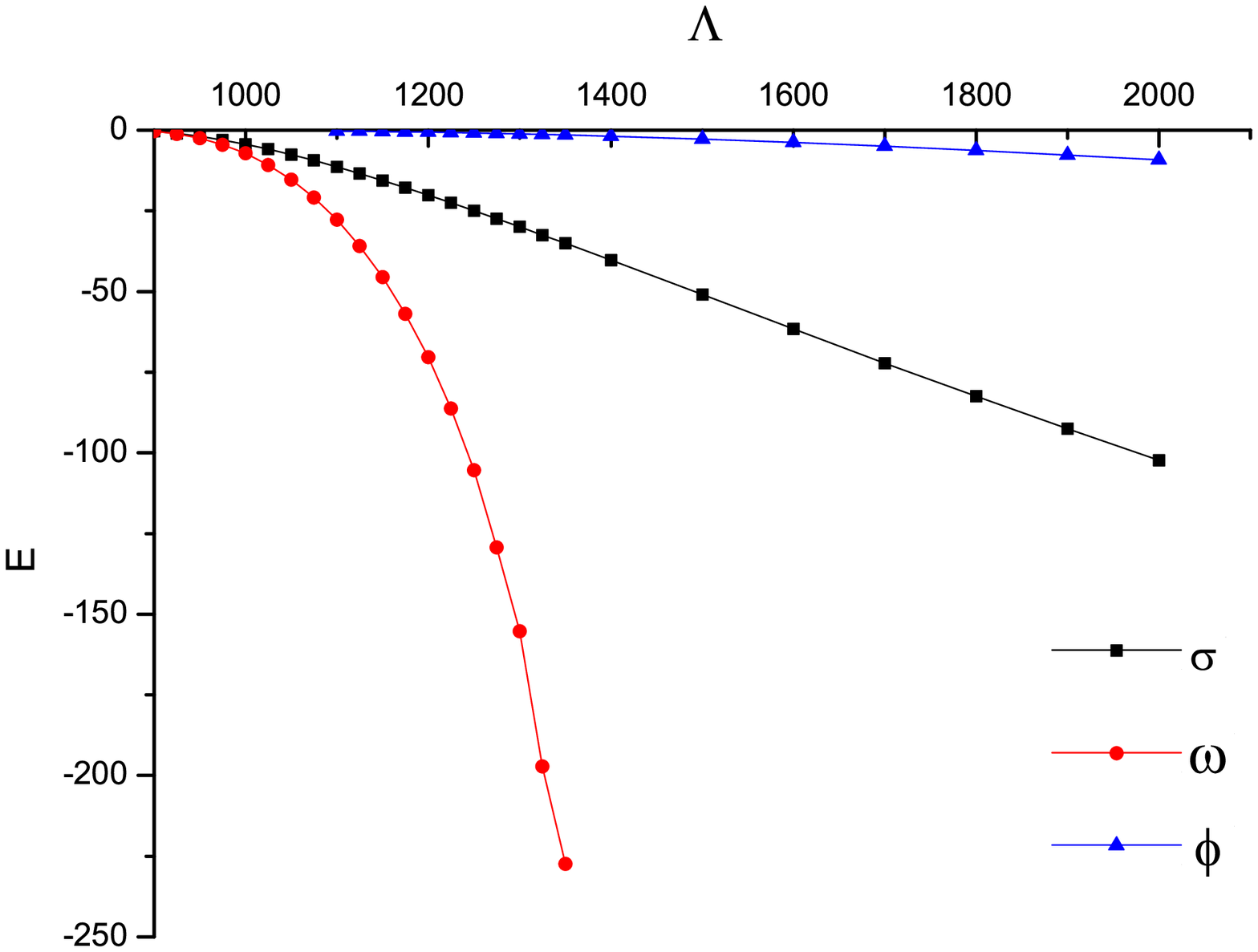}}
    \rotatebox{0}{\includegraphics*[width=0.47\textwidth]{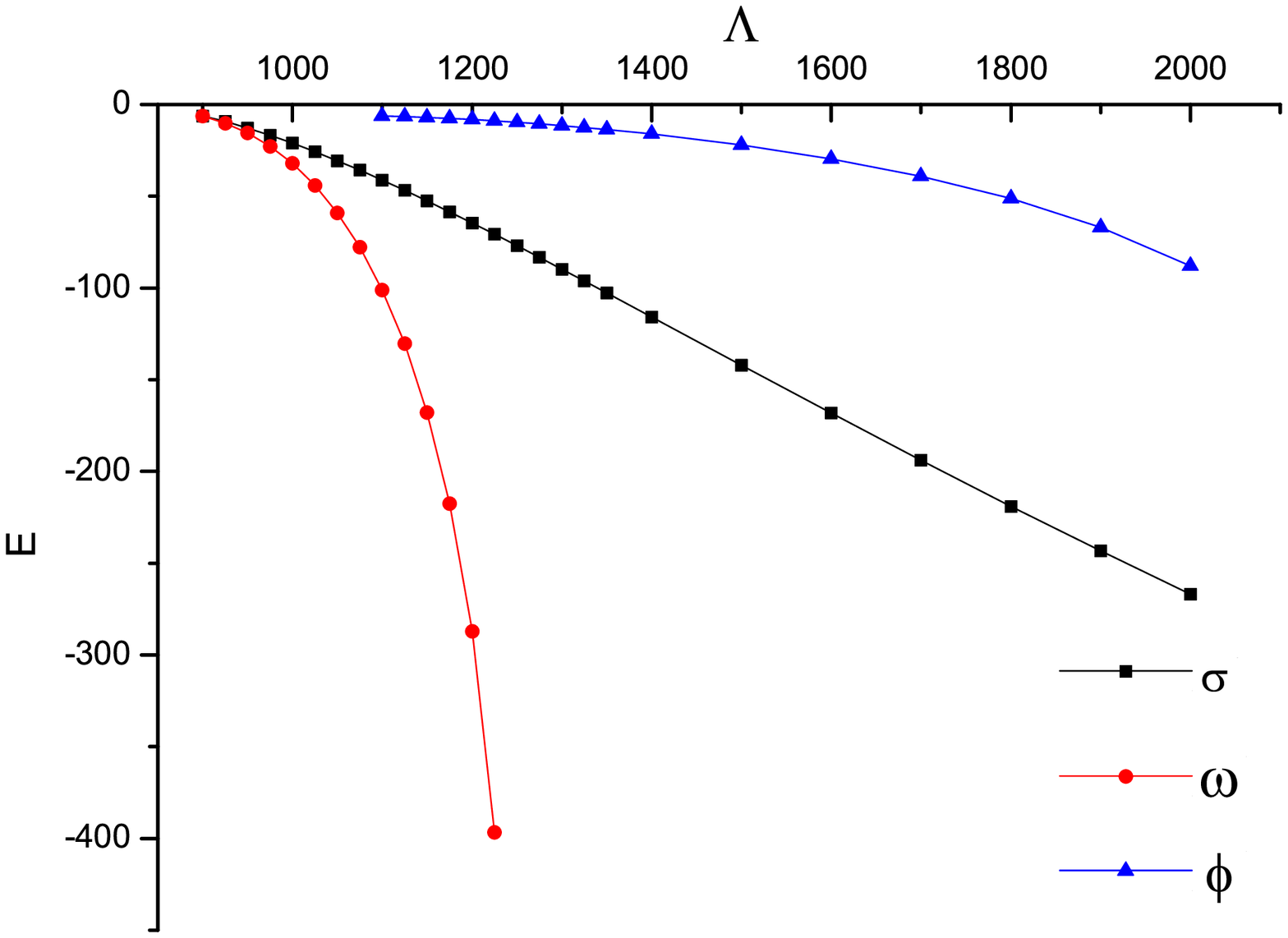}}
    \caption{(Color online) The dependence of the binding energy on $\Lambda_{\sigma}$, $\Lambda{\omega}$ and $\Lambda_{\phi}$ for the states $^1S_0$ (left) and $^3S_1$ (right). The cutoff parameters for the $\eta$-exchange and $\eta^{'}$-exchange are fixed to be $\Lambda_{\eta}=1900$ MeV and $\Lambda_{\eta^{'}}=2000$ MeV respectively.}
    \label{cutofffig}
  \end{center}
\end{figure}

Meanwhile, we also perform an investigation of the dependence of the
binding energies on $\Lambda_{\sigma}$, $\Lambda_{\omega}$ and
$\Lambda_{\phi}$. During our study, we change one of the above three
parameters in its proper range while keeping the other two to be
their lowest value. The cutoff parameter for the individual meson
should be larger than mass of the exchanged meson, so the lowest
values of the cutoff parameters can be taken as
$\Lambda_{\sigma}=900$ MeV, $\Lambda_{\omega}=900$ MeV, and
$\Lambda_{\phi}=1100$ MeV. We plot the variation of the binding
energy with individual cutoff parameter in Fig.~\ref{cutofffig}.

From the curves of Fig~\ref{cutofffig}, we see that the binding
energies of both $^1S_0$ and $^3S_1$ depend most sensitively on
$\Lambda_{\omega}$ and least sensitively on $\Lambda_{\phi}$. Since
the solutions changes dramatically when the cutoff parameter of the
$\omega$-exchange increases, it seems that we should to take
$\Lambda_{\omega}<1000$ MeV. On the other hand, the binding energy
changes very slowly with $\Lambda_{\phi}$, so we fix the cutoff
parameter of the $\phi$-exchange between $1700$ MeV and $1800$ MeV.
Given the sigma is lighter than the omega, we take
$\Lambda_{\sigma}=900$ MeV. Based on these analysis, we tabulated
the numerical results in Table~\ref{newset}. From
Table~\ref{newset}, we can see that when the cutoff parameters are
taken as $\Lambda_{\sigma}=900$ MeV,
$\Lambda_{\omega}\subset(900\sim1000~\mbox{MeV})$ and
$\Lambda_{\phi}\subset(1700\sim1800~\mbox{MeV})$, we obtain a
loosely bound state of $\Lambda\bar{\Lambda}(^1S_0)$, with binding
energy around $-7.624\sim-13.290$ MeV. The state
$\Lambda\bar{\Lambda}(^3S_1)$ is also a loosely bound state, with
slightly larger binding energy around $-50.389\sim-82.744$ MeV.

\begin{table}
\renewcommand{\arraystretch}{1.5}
\caption{The binding energies of the states
$\Lambda\bar{\Lambda}(^1S_0)$ and $\Lambda\bar{\Lambda}(^3S_1)$ with
the $\Lambda_{\sigma}=900$ MeV,
$900~\mbox{MeV}<\Lambda_{\omega}<1000~\mbox{MeV}$ and
$1700~\mbox{MeV}<\Lambda_{\phi}<1800~\mbox{MeV}$. The cutoff
parameters for the $\eta$-exchange and $\eta^{'}$-exchange are fixed
to be $\Lambda_{\eta}=1900$ MeV and $\Lambda_{\eta^{'}}=2000$ MeV
respectively.} \label{newset}
\begin{center}
\begin{tabular*}{8.5cm}{@{\extracolsep{\fill}}ccccc}
\hline \hline
                   &                          &                       &
\multicolumn{2}{c}{E (MeV)}\\
 $\Lambda_{\sigma}$ (MeV)&$\Lambda_{\omega}$ (MeV)&$\Lambda_{\phi}$ (MeV)& $^{3}S_{1}$        & $^{1}S_{0}$\\
\hline
 900               & 925                      & 1700                 & $-50.389$                   & $-7.624$\\
 900               & 925                      & 1750                &
$-57.221$                   & $-8.441$\\
 900               & 925                      & 1800                & $-64.950$                   & $-9.290$\\
 900               & 950                      & 1700                 & $-64.891$                   &$-11.153$\\
900               & 950                      & 1750                 & $-73.241$                   &$-12.203$\\
 900               & 950                      & 1800                 & $-82.744$                  &$-13.290$\\
\hline\hline
\end{tabular*}
\end{center}
\end{table}

The threshold of $\Lambda\bar{\Lambda}$ is 2231.3 MeV. If $Y(2175)$
and $\eta(2225)$ are regarded as bound states of
$\Lambda\bar{\Lambda}(^3S_1)$ and $\Lambda\bar{\Lambda}(^1S_0)$, the
binding energies should be $-56.37$ MeV and $-6.37$ MeV
respectively, both of which roughly lie in the range of the our
results, $-50.4\sim-82.7$ MeV for state
$\Lambda\bar{\Lambda}(^3S_1)$ and $-7.6\sim -13.3$ Mev for states
$\Lambda\bar{\Lambda}(^1S_0)$. From Table~\ref{newset}, we can also
tell that the difference between the binding energy of the state
$^3S_1$ and that of the state $^1S_0$ increases slowly when the
attractive forces coming from the $\omega$-exchange and
$\phi$-exchange increase. Besides, we also notice that the
difference of the binding energy between these two states is
$43\sim69$ Mev, which is consistent with the difference between the
thresholds of $Y(2175)$ and that of $\eta(2225)$. This prominent
feature seems to indicate that $Y(2175)$ and $\eta(2225)$ might be
regarded as the bound states of $\Lambda\bar{\Lambda}(^3S_1)$ and
$\Lambda\bar{\Lambda}(^1S_0)$ respectively.

Besides, we also perform a study of the hidden-charm partner of
$\Lambda\bar{\Lambda}$ because of the similarity of the
$\Lambda\bar{\Lambda}$ and $\Lambda_c\bar{\Lambda_c}$ systems.
Actually, in Ref.~\cite{Lee:2011rka} the authors have studied the
baryonium $\Lambda_c\bar{\Lambda}_c$. However, they omit the term
related to $\mathcal{F}_4$ (see Eq.~\ref{transform-4}) which also
appears in the study of the deuteron~\cite{Machleidt1987}. We first
reproduce their results with our program and then focus on the
contribution of the terms related to $\mathcal{F}_4$ in the
formation of the bound states of $\Lambda_c\bar{\Lambda}_c$ and
$\Lambda\bar{\Lambda}$. We summarize our results in
Tables~\ref{lamdac},~\ref{1S0} and \ref{3S1}.

Our result indicates that the terms related to $\mathcal{F}_4$ have
tiny influence on the bound state of the hidden-charm
$\Lambda_c\bar{\Lambda}_c$, see Table~\ref{lamdac}. These terms also
change the binding of the bound state of
$\Lambda\bar{\Lambda}(^1S_0)$ very little, see Table~\ref{1S0}.
However, these terms can deepen the binding of the state
$\Lambda\bar{\Lambda}(^3S_1)$ significantly when binding energy of
the state $\Lambda\bar{\Lambda}(^3S_1)$ reaches tens of MeV. For
example, when the cutoff parameter is fixed to be 1100 MeV, the
binding energy of the state $\Lambda\bar{\Lambda}(^3S_1)$ is
$-57.974$ MeV without the terms $\mathcal{F}_4$-related terms.
However, it changes into $-101.066$ with the $\mathcal{F}_4$-related
terms included, see Table~\ref{3S1}.

\begin{table}
\renewcommand{\arraystretch}{1.5}
\caption{The contribution of the term $\mathcal{F}_4$ in forming the
bound state of $\Lambda_c\bar{\Lambda}_c(^1S_0)$. $\Lambda$ is the
cutoff parameter. The result without $\mathcal{F}_4$ (original)
comes from Ref.~\cite{Lee:2011rka}.} \label{lamdac}
\begin{center}
\begin{tabular*}{8.5cm}{@{\extracolsep{\fill}}ccc}
\hline \hline
         ~             &\multicolumn{2}{c}{E (MeV)}\\
$\Lambda$(MeV)         &  original                   &
$\mathcal{F}_{4}$ added
\\ \hline
          890          &     $-2.80$                 & $-2.88$  \\
 900                   &     $-4.61$                 & $-4.75$  \\
1000                   &     $-49.72$                & $-53.20$ \\
1100                   &     $-142.19$               & $-160.115$ \\
\hline\hline \label{lamdac}
\end{tabular*}
\end{center}
\end{table}

\begin{table}
\renewcommand{\arraystretch}{1.5}
\caption{The contribution of the terms related to $\mathcal{F}_4$ in
forming the bound state of $\Lambda\bar{\Lambda}(^1S_0)$. Here, we
adopt the same value for the cutoff parameters of all the exchanged
mesons.} \label{1S0}
\begin{center}
\begin{tabular*}{8.5cm}{@{\extracolsep{\fill}}ccc}
\hline \hline
         ~          &\multicolumn{2}{c}{E (MeV)}\\
 $\Lambda$ (MeV)    & without $\mathcal{F}_{4}$& with $\mathcal{F}_{4}$\\
 \hline
 900                & $-0.317$                 & $-0.274$              \\
 925                & $-1.150$                 & $-1.140$              \\
 950                & $-2.409$                 & $-2.492$              \\
 975                & $-4.192$                 & $-4.474$              \\
1000                & $-6.568$                 & $-7.203$              \\
1025                & $-9.580$                 & $-10.783$             \\
1050                & $-13.261$                & $-15.316$             \\
1075                & $-17.632$                & $-20.914$             \\
1100                & $-22.711$                & $-27.703$             \\
\hline\hline
\end{tabular*}
\end{center}
\end{table}

\begin{table}
\renewcommand{\arraystretch}{1.5}
\caption{The contribution of the terms related to $\mathcal{F}_4$ in
forming the bound state of $\Lambda\bar{\Lambda}(^3S_1)$. Here, we
adopt the same value for the cutoff parameters of all the exchanged
mesons.} \label{3S1}
\begin{center}
\begin{tabular*}{8.5cm}{@{\extracolsep{\fill}}ccc}
\hline \hline
      ~               &\multicolumn{2}{c}{E (MeV)} \\
$\Lambda$ (MeV)       & without $\mathcal{F}_{4}$ & with $\mathcal{F}_{4}$ \\
\hline
        900           & $-6.549$                  & $-6.258$               \\
        925           & $-9.728$                  & $-10.125$              \\
        950           & $-13.822$                 & $-15.52$               \\
        975           & $-18.868$                 & $-22.742$              \\
       1000           & $-24.873$                 & $-32.126$              \\
       1025           & $-31.823$                 & $-44.065$              \\
       1050           & $-39.687$                 & $-59.054$              \\
       1075           & $-48.421$                 & $-77.748$              \\
       1100           & $-57.974$                 & $-101.066$             \\
\hline\hline \label{3S1}
\end{tabular*}
\end{center}
\end{table}

In Ref.~\cite{Lee:2011rka}, the authors also studied the
spin-triplet $\Lambda_c\bar{\Lambda}_c$ where the S-D mixing effect
may be important. In their study, they related the coupling
constants of Lambda-Lambda-meson to those of nucleon-nucleon-meson
via the quark model. Since there exists $\mbox{SU}(3)-flavor$
symmetry breaking of the coupling constants of
nucleon-nucleon-meson, the authors adopted the values
$f_{\omega\Lambda_c\Lambda_c}=-g_{\omega\Lambda_c\Lambda_c}$ which
leads to the vanishing S-D mixing. In the present case we take
$f_{\omega\Lambda_c\Lambda_c}=0$ because $f_{\omega NN}=0$. Now we
revisit the spin-triplet $\Lambda_c\bar{\Lambda}_c$ system. We
mainly focus on the effect of the S-D mixing in forming the bound
state of $\Lambda_c\bar{\Lambda}_c$ with spin-triplet. We summarize
our results in Table~\ref{lamdac-vector-coup}. Our result indicates
that the effect of the S-D mixing in the formation of the bound
state of $\Lambda_c\bar{\Lambda}$ with spin-triplet is quite small.
For example, when we set the cutoff parameter to be 900 MeV, the
binding energy without the S-D mixing is $-4.61$ MeV. When we add
the S-D mixing, the binding energy is $-4.40$ MeV with the same
cutoff parameter, see Table~\ref{lamdac-vector-coup}.

\begin{table}
\renewcommand{\arraystretch}{1.5}
\caption{The contribution of the S-D mixing in forming the bound
state of $\Lambda_c\bar{\Lambda}_c$. Here, we adopt the coupling
constant for the $\omega$-exchange as
$f_{\omega\Lambda_c\Lambda_c}=0$. The results of ``original" are
taken from Ref.~\cite{Lee:2011rka}.} \label{lamdac-vector-coup}
\begin{center}
\begin{tabular*}{8cm}{@{\extracolsep{\fill}}cccc}
\hline \hline
     ~                &\multicolumn{3}{c}{E (MeV)}\\
$\Lambda$ (MeV)       & original          &  $^1S_0$    & $^3S_1- ^3D_1$ \\
\hline
 890                  & $-2.80$           &  $-2.80$    & $-2.66$       \\
 900                  & $-4.61$           &  $-4.61$    & $-4.40$       \\
1000                  & $-49.72$          &  $-49.72$   & $-46.50$      \\
1100                  & $-142.19$         &  $-142.19$  & $-130.17$     \\
\hline\hline
\end{tabular*}
\end{center}
\end{table}

\section{Summary and Discussion}\label{summary}

In the present work, we have used the one-boson-exchange potential
(OBEP) model, which works very well in describing the deuteron, to
study the system of $\Lambda\bar{\Lambda}$ with quantum numbers
$J^{PC}=1^{--}$ and $0^{-+}$. We have included the contributions of
the pseudoscalar $\eta$ and $\eta^{'}$ exchanges, the scalar
$\sigma$-exchange and the vector $\omega$ and $\phi$ exchanges.
Since the reasonable range of the cutoff parameter in the study of
the deuteron is $800\sim1500$ MeV, we take the range as
$900\sim2000$ MeV which is wide enough to study the dependence of
the binding solutions on the cutoff parameter. We follow the rule
that the heavier the meson is, the larger the cutoff is and that the
cutoff parameter should be larger than the mass of the corresponding
exchanged mesons.

Our results indicate that both the $\eta$-exchange and the
$\eta^{'}$-exchange provide repulsive force for the state $^1S_0$
but attractive force for the state $^3S_1$. The $\sigma$-exchange
provides attractive force for both of these two states. If we fix
the cutoff parameters for the $\eta$-exchange and the
$\eta^{'}$-exchange to be $\Lambda_{\eta}=1900$ MeV and
$\Lambda_{\eta^{'}}=2000$ MeV respectively, we find the binding
solutions for both of the two states depend most sensitively on
$\Lambda_{\omega}$ and least sensitively on $\Lambda_{\phi}$. We
also find that the binding of the state $^1S_0$ is shallower than
that of $^3S_1$ with the same cutoff parameter.

When we fix $\Lambda_{\eta}=1900$ MeV, $\Lambda_{\eta}^{'}=2000$ MeV
and $\Lambda_{\sigma}=900$ MeV and tune $\Lambda_{\omega}$ between
$900$ MeV and $1000$ MeV and $\Lambda_{\phi}$ between $1700$ and
$1800$ MeV, we obtain bound states for both $^1S_0$ and $^3S_1$. The
binding energies are $-7.6\sim-11.3$ MeV and $-50.4\sim-82.7$ MeV
respectively. Assuming $Y(2175)$ and $\eta(2225)$ are bound states
of $\Lambda\bar{\Lambda}(^3S_1)$ and $\Lambda\bar{\Lambda}(^1S_0)$,
the binding energies should be $-56.37$ MeV and $-6.37$ MeV
respectively which lie in the ranges of our results,
$-50.4\sim-82.7$ MeV and $-7.6\sim -11.3$ MeV. Most importantly, we
also notice that the difference of the binding energies between the
state $^3S_1$ and $^1S_0$ is $43\sim69$ MeV which is consistent with
the difference between the masses of $Y(2175)$ and $\eta(2225)$. Our
present calculation suggests that $Y(2175)$ and $\eta(2225)$ may be
the bound states of $\Lambda\bar{\Lambda}(^3S_1)$ and
$\Lambda\bar{\Lambda}(^1S_0)$. The study of their decay patterns
within the same framework will be very helpful. In fact, there is
some evidence for the $\Lambda\bar{\Lambda}$ near-threshold
enhancement in the
$J/\psi\to\gamma\Lambda\bar{\Lambda}$~\cite{Daijp}, which may be due
to the $\eta(2225)$.

Because of the similarity of $\Lambda\bar{\Lambda}$ and
$\Lambda_c\bar{\Lambda}_c$, we also perform a study of the
hidden-charm partner of $\Lambda\bar{\Lambda}$. Given that the
authors in Ref.~~\cite{Lee:2011rka} have studied the baryonium of
$\Lambda_c\bar{\Lambda}_c$. We first confirm their results and then
focus on the contribution of the terms related to $\mathcal{F}_4$ in
forming the bound states of $\Lambda\bar{\Lambda}$ and
$\Lambda_c\bar{\Lambda}_c$. From our results, we find that the
contribution of the terms related to $\mathcal{F}_4$ is small for
the system $\Lambda_c\bar{\Lambda}_c$. The case of the state
$\Lambda\bar{\Lambda}(^1S_0)$ is similar. However, for the
spin-triplet state of $\Lambda\bar{\Lambda}$, the
$\mathcal{F}_4$-related terms change the binding energy
significantly when the binding energy is around tens of MeV. We also
find the S-D mixing provides quite small contributions in the
formation of the spin-triplet state of $\Lambda_c\Lambda_c$.

\section{acknowledgement}
We thank Jian-Ping Dai and Bo-Chao Liu for useful discussions. This
work is supported by the National Natural Science Foundation of
China under Grant 11035006, 11121092, 11261130311 (CRC110 by DFG and
NSFC), the Chinese Academy of Sciences under Project No.KJCX2-EW-N01
and the Ministry of Science and Technology of China (2009CB825200).


\end{document}